\documentclass[spanish,english,referee]{aa}

\usepackage{txfonts,epsfig,graphicx,natbib}

\bibpunct{(}{)}{;}{a}{}{,}    



\author{E. Retana-Montenegro and 
F. Frutos-Alfaro \\ 
Escuela de F\'isica \\
Universidad de Costa Rica \\
San Pedro 11501, Costa Rica}

\title{The lensing properties of the Einasto profile}

\date{\today}

\institute{Escuela de Física, Universidad de Costa Rica, San Pedro 11501, Costa
Rica \\
\email{edwin@fisica.ucr.ac.cr}\\
}

\begin{document}
\onecolumn{%
\vspace*{4ex}
\begin{center}
  {\Large \bf The lensing properties of the Einasto profile}\\[4ex]   
  {\large \bf E. Retana-Montenegro and F. Frutos-Alfaro }\\[4ex]
  \begin{minipage}[t]{16cm} 
             Escuela de F\'isica,
             Universidad de Costa Rica, San Pedro 11501,
             Costa Rica\\
             
\email{edwin@fisica.ucr.ac.cr}

{\bf Abstract.}
In recent high resolution N-body CDM simulations, it has been had found that nonsingular
three-parameter models, e.g. the Einasto profile has a better performance
better than the singular two-parameter models, e.g. the Navarro, Frenk
and White in the fitting of a wide range of dark matter halos. A problem
with this profile is that the surface mass density is non-analytical
for general values of the Einasto index. Therefore, its other lensing
properties have the same problem. We obtain an exact analytical expression
for the surface mass density of the Einasto profile in terms of the
Fox H-function for \emph{all} values of the Einasto index. With the
idea of facilitate the use of the Einasto profile in lensing studies,
we calculate the surface mass density, deflection angle, lens equation,
deflection potential, magnification, shear and critical curves of
the Einasto profile in terms of the Meijer G-function for \emph{all}
rational values of the Einasto index. The Meijer G-function have been
implemented in several commercial and open-source computer algebra
systems, thus the use of the lensing properties of the Einasto profile
in strong and weak lensing studies is straighforward. We also compare
the S\'ersic and Einasto surface mass densities profiles and found 
differences between them. This implies that the lensing properties are
not equal for both profiles.
     
\vspace*{2ex}
\end{minipage}
\end{center}
}

             

\section{Introduction}

\noindent
The Cold Dark Matter (CDM) theory has become the standard theory of 
cosmological structural formation. Its variant the $\Lambda CMD$ 
with $\left(\Omega_{m}, \, \Omega_{\Lambda}\right)=\left(0.3, \, 0.7\right)$
seems to be in agreement with the observations on cluster-sized scales
(\citealt{2003NuPhS.124....3P}). On galaxy/sub-galaxy scales has
several problems, such as the discrepancy between observations and
the results of numerical simulations. The high resolution observations
of rotation curves of low surface brightness (LSM) and dark matter
dominated dwarf galaxies \citep{2001ApJ...552L..23D,
2001MNRAS.325.1017V,2003ApJ...583..732S,2003MNRAS.340...12W,
2004MNRAS.353L..17D,2005ApJ...634L.145G,2005ApJ...621..757S,
2007MNRAS.375..199G,2010NewA...15...89B}
favor density profiles with a flat central core (e.g. 
\citealt{1995ApJ...447L..25B,2000ApJ...537L...9S,
2004MNRAS.351..903G,2009RAA.....9.1173L}).
In contrast N-body CDM simulations predict a two parameter functional
form for the density profiles with too high densities (cusps) in the
galatic center \citep{1996ApJ...462..563N,1997ApJ...490..493N,
1999MNRAS.310.1147M}.
This discrepancy is called the cusp-core problem.

\noindent
Gravitational lensing is one of the most powerful tools in observational
cosmology for probing the distribution of matter of collapsed objects
like galaxies and clusters in strong \citep{1989MNRAS.238...43K,
1994AJ....108.1156W,1996A&A...313..697B,1998ApJ...495..609C,
2000ApJ...535..692K,2001ApJ...549L..25K,2002ApJ...574L.129S,
2002ApJ...575L...1K,2003ApJ...582...17K,2004ApJ...612..660K,
2008A&A...489...23L,2009A&A...507...35A,2011MNRAS.413.1753Z,
2011arXiv1107.2649Z} and weak \citep{1993ApJ...404..441K,1999ARA&A..37..127M,
2001PhR...340..291B,2004ApJ...606...67H,2006ApJ...648L.109C,
2007ApJ...668..806M,2009ApJ...704..672J,2011A&A...529A..93H} regimes. 
In order to obtain the fit to the observational data in strong
and weak lensing studies the most accurate density profile must be
used. The first step before using a profile in lensing studies is
to investigate the lensing properties of the profile.

\noindent
Recently N-body CDM simulations \citep{2004MNRAS.349.1039N,2006AJ....132.2685M,
2008MNRAS.387..536G,2008MNRAS.388....2H,2009MNRAS.398L..21S,
2010MNRAS.402...21N}
have found that the three-parameter profiles fit better to a wide
range of dark matter halos. One of this profiles is the 
\citet{1965TIAAA.51..87}
profile, a 3D version of the 2D \citet{1968adga.book.....S} model
used to described the surface brightness of galaxies. The S\'ersic profile
can be written as: 

\begin{equation}
\Sigma_{S}\left(R\right)=\Upsilon I_{e}\exp\left\{ 
-b_{n}\left[\left(\frac{R}{R_{e}}\right)^{1/n}-1\right]\right\} , 
\label{eq:Sersic-surface-mass-density}
\end{equation}

\noindent 
where $R$ is the distance in the sky plane, $n$ the S\'ersic 
index,$\Upsilon$ the mass-to-light ratio, $I_{e}$ the luminosity
density at the effective radius $R_{e}$, $b_{n}$ is a function of
$n$ that can be determined from the condition that the luminosity
inside $R_{E}$ equals half of the total luminosity, for example 
\citet{1997A&A...321..111P} found $b_{n}=2n-0.3333+0.009876/n$.

The Einasto profile is given by:

\noindent 
\begin{equation}
\rho\left(r\right)=\rho_{E}\exp\left(-d_{\alpha}\left[\left(\frac{r}{r_{E}}
\right)^{\alpha}-1\right]\right)\label{eq:01}
\end{equation}

\noindent 
where $r$ is the spatial radius, $\alpha$ is the Einasto
index that determines the shape of the profile, $d_{\alpha}$ is a
function of $\alpha$ which allows the calculation of the density
$\rho_{E}$ inside an effective radius $r_{E}$. In the context of
dark matter halos this can expressed as:

\noindent 
\begin{equation}
\rho\left(r\right)=\rho_{-2}\exp\left(-\frac{2}{\alpha}\left[
\left(\frac{r}{r_{-2}}\right)^{\alpha}-1\right]\right)\label{eq:02}
\end{equation}

\noindent 
where $\rho_{-2}$ and $r_{-2}$ are the density and radius
at which $\rho\left(r\right)\propto r^{-2}$. Both radius and densities
are related by $\rho_{-2}=\rho_{E}\exp\left(2/\alpha-d_{\alpha}\right)$
and $r_{-2}=r_{E}\left(\alpha d_{\alpha}/2\right)^{\alpha}$. First
\citet{2004MNRAS.349.1039N} found that for haloes with masses from
dwarfs to clusters $0.12\lesssim\alpha\lesssim0.22$ with an average
value of $\alpha=0.17$. \citet{2008MNRAS.388....2H} and 
\citet{2008MNRAS.387..536G}
found that there are trends for $\alpha_{E}$ to increase with mass
and redshift, with $\alpha\sim0.17$ for galaxy and $\alpha\sim0.23$
for cluster-sized haloes in the Millennium Simulation (MS) 
\citep{2005Natur.435..629S}. \citet{2010MNRAS.402...21N} found similar results 
for galaxy-sized haloes in the Aquarius simulation 
\citep{2008MNRAS.391.1685S}. Also, \citet{2008MNRAS.387..536G} found that 
$\alpha\sim0.3$ for the most massive haloes of the MS.

\noindent
A problem with the Einasto profile is that in order to study its lensing
properties numerical methods must be used because analytical expressions
for this properties are not available. A semi-analytical approximation
for the projected 2D projection of the Einasto was obtained by 
\citet{2010MNRAS.405..340D}. A recent work done by \citet{2011A&A...525A.136B} 
demonstrated that is possible to write analytical expressions for the deprojected 
S\'ersic model in terms of the Meijer G-function using Mellin integral 
transforms.

\noindent
In this paper, we present analytical expressions for the lensing properties
of the Einasto profile. We apply the Mellin-transform method to derive
an analytical expression for the projected surface mass density, the deflection
angle, the lens equation, the deflection potential, the shear, the
tangential and critical curves for all values of the Einasto
index $\alpha$ in terms of the Meijer G-function. This function
can be automatically handed by numerical routines implemented in computer
algebra systems (CAS) such as the commercial $Mathematica^{\textregistered}$
and $Maple^{\textregistered}$ and the free open-source $Sage$
and in the Python library $mpmath$.

\noindent
This paper is organized as follows. In Section \ref{sec:02} we derive
the surface mass density of the Einasto profile in terms of the Fox H-function 
and the Meijer G-function for values
$\alpha=\frac{1}{n}$ and $\alpha=\frac{2}{n}$ with $n$ integer,
and rational values of the Einasto index. In Section \ref{sec:03}
we evaluate the total mass enclosed by this class of models using
the density profile and the surface mass density obtained in the previous
section. In Section \ref{sec:04} we use the result for the projected
surface mass density to calculate the deflection angle, the lens equation
and deflection potential for a spherically symmetric lens described
by the Einasto profile in terms of the Meijer G-function. In
Section \ref{sec:05} we derive expressions for the magnification,
shear and the critical curves of the Einasto profile. We summary our
conclusions in Section \ref{sec:06}. We give an brief description
of the Mellin transform-method in Appendix \ref{sec:A}. In Appendix
\ref{sec:B} we formulate all the properties of the Fox H-functions and 
Meijer G-functions that are used in this work.

\section{Analytical expression for the surface mass density of the Einasto
profile\label{sec:02}}

\noindent
The projected surface mass density of a spherically symmetric lens is given
by integrating along the line of sight the 3D density profile:

\begin{equation}
\Sigma\left(\xi\right)=\int_{-\infty}^{+\infty}\rho\left(\xi,r\right)dz ,
\label{eq:03}
\end{equation}

\noindent
where $\xi$ is the radius measure from the centre of the lens and
$r=\sqrt{\xi^{2}+z^{2}}$. This expression can also be written as
an Abel integral \citep{1987gady.book.....B}:

\begin{equation}
\Sigma\left(\xi\right)=2\int_{\xi}^{\infty}\frac{\rho\left(r\right)rdr}
{\sqrt{r^{2}-\xi^{2}}}\label{eq:04}
\end{equation}

\noindent 
By inserting equation (\ref{eq:02}) into the above expression 

\begin{equation}
\Sigma\left(x\right)=2\rho_{-2}r_{-2}e^{\frac{2}{\alpha}}\int_{x}^{\infty}
\frac{\exp\left(\frac{-2s^{\alpha}}{\alpha}\right)sds}{\sqrt{s^{2}-x^{2}}}
\label{eq:05}
\end{equation}

\noindent 
having introduced the quantities $x=\xi/r_{-2}$ and $s=r/r_{-2}$.

\noindent 
The integral (\ref{eq:05}) can not be expressed in terms of ordinary
functions for all the values of $\alpha_{E}$. However, using the
Mellin-transform method \citep{0853125287,1996MER.16..05,159829184X}
is possible the exact calculation of one dimensional definite integrals.
The most powerful feature of this method is that the result is a Mellin-Barnes
integral. This integral for a certain combination of coefficients
is the integral representation of a Fox H-function or a Meijer
G-function (for details see Appendix \ref{sec:A}) .

\noindent 
Using the Mellin-transform method with the integral (\ref{eq:05}),
with $z=1$ and the functions:

\begin{equation}
f\left(s\right)=2\rho_{-2}r_{-2}e^{\frac{2}{\alpha}}\exp\left(
\frac{-2s^{\alpha}}{\alpha}\right)\label{eq:06}
\end{equation}

\begin{equation}
g\left(s\right)=\left\{ \begin{array}{cc}
{\displaystyle \frac{1}{s\sqrt{1-\left(sx\right)^{2}}}} & 0\leq s\leq x^{-1}\\
0 & elsewhere
\end{array}\right.\label{eq:07}
\end{equation}

\noindent 
and its Mellin transforms:

\begin{equation}
\left\{ \mathcal{M}f\right\} \left(u\right)=2\rho_{-2}r_{-2}
e^{\frac{2}{\alpha}}\alpha^{-1}\left(\frac{2}{\alpha}\right)^{-u/\alpha}
\Gamma\left(\frac{u}{\alpha}\right)\label{eq:08}
\end{equation}

\begin{equation}
\left\{\mathcal{M}g\right\} \left(u\right)=\frac{\sqrt{\pi}}{4x^{u-1}}\frac{
\Gamma\left(\frac{u-1}{2}\right)u}{\Gamma\left(1+\frac{u}{_{2}}\right)} 
\label{eq:09}
\end{equation}

\noindent 
Combining equations (\ref{eq:08}), (\ref{eq:09}) and 
(\ref{eq:central_feature_mellin}) with $u=2y$ and $m=1/\alpha$ yields:

\begin{equation}
\Sigma\left(x\right)=\sqrt{\pi}\rho_{-2}r_{-2}e^{2m}\: x\:\frac{1}{2\pi i}
\int_{_{\mathit{C}}}\frac{\Gamma\left(-\frac{1}{2}+y\right)
\Gamma\left(1+2my\right)}{\Gamma\left(1+y\right)}\left[
\left(2m\right)^{2m}x^{2}\right]^{-y}dy \label{eq:Einasto-mellin-transform}
\end{equation}

\noindent 
Comparing the last equation with (\ref{eq:Fox-H-1}) is possible to obtain an 
analytical expression in terms of the Fox H-function for the surface mass 
density of the Einasto profile:

\begin{equation}
\Sigma\left(x\right)=\sqrt{\pi}\rho_{-2}r_{-2}e^{\frac{2}{\alpha}}\: x\: 
H_{1,2}^{2,0}\left[\begin{array}{c}
\left(1,\,1\right)\\
\left(1,\,\frac{2}{\alpha}\right),\left(-\frac{1}{2},\,1\right)
\end{array}\biggr|\;\left(\frac{2}{\alpha}\right)^{\frac{2}{\alpha}} 
x^{2}\right]
\end{equation}

\noindent 
Writting the surface mass density as a Fox H-function
has an inconvenient. The Fox H-function despite having a great
potential for analytical work in Mathematics, sciences and engineering
no numerical routines has been implemented yet. We prefer to describe
the lensing properties of the Einasto profile in terms of analytical
functions that have numerical routines already implemented to facilitate
its use in strong and weak lensing studies.

\subsection{Surface mass density of the Einasto profile as a Meijer G-function}

\noindent 
The Meijer G-function meets the requirement pointed out before.
A list of the relevant properties of the Meijer G-function
can be found in Appendix \ref{sec:B}. We can use this function to
write expressions in analytical form for most of the lensing properties
of the Einasto profile. The Meijer G-function had been
implemented in several commercial and free available CAS. This means
that using the Meijer G-function in lensing studies
is just as simple as use other special functions like Hypergeometric,
Gamma and Bessel functions for example.

\noindent 
Using a similar procedure to the one used by \citet{2011A&A...525A.136B}
to obtain an analytical expression for the luminosity density in terms
of the Meijer G-function for all rational values of the S\'ersic index
we proceed to do the same to derive an expression for the surface
mass density of the Einasto profile for all values of the Einasto
index.

\subsubsection{Einasto index with values $\alpha=\frac{1}{n}$ and 
$\alpha=\frac{2}{n}$ with $n$ integer}

\noindent 
The equation (\ref{eq:Einasto-mellin-transform}) can be written in terms 
of the Meijer G-function for the Einasto index with values 
$\alpha=\frac{1}{n}$ and $\alpha=\frac{2}{n}$ with $n$ integer.
But first, one substitution is required using the Gauss Multiplication
formula \citep{1970hmfw.book.....A}:

\begin{equation}
\prod_{j=0}^{N-1}\Gamma\left(z^{\prime}+\frac{j}{N}\right)=
\left(2\pi\right)^{\frac{N-1}{2}}N^{\frac{1}{2}-Nz^{\prime}}
\Gamma\left(Nz^{\prime}\right) , \label{eq:gauss_mult_form}
\end{equation}

\noindent 
with $z^{\prime}=z/N$, $N=2m$ and $z=2my$, we get:

\begin{equation}
\Gamma\left(1+2my\right)=\left(2m\right)^{\frac{1}{2}+2my}
\left(2\pi\right)^{\frac{1}{2}-m}\Gamma\left(1+y\right)
\prod_{j=1}^{2m-1}\Gamma\left(\frac{j}{2m}+y\right) \label{eq:primer-gamma}
\end{equation}

\noindent 
Substituting the last equation into (\ref{eq:Einasto-mellin-transform}),
we obtain:

\begin{equation}
\Sigma\left(x\right)=\frac{\rho_{-2}r_{-2}\sqrt{m}}{\left(2\pi\right)^{m-1}}
e^{2m}x\:\frac{1}{2\pi i}\int_{_{\mathit{C}}}\Gamma\left(-\frac{1}{2}+y\right)
\prod_{j=1}^{2m-1}\Gamma\left(\frac{j}{2m}+y\right)\left[x^{2}\right]^{-y}dy
\label{eq:Mellin-transform-m-interger}
\end{equation}

\noindent 
Comparing with the integral representation of the Meijer G-function 
(\ref{eq:Meijer-G}) we found an analytical expression
for the surface mass density of the Einasto profile:

\begin{equation}
\Sigma\left(x\right)=\frac{\rho_{-2}r_{-2}e^{\frac{2}{\alpha}}}{
\left(2\pi\right)^{\frac{1}{\alpha}-1}\sqrt{\alpha}}\: x\: 
G_{0,\frac{2}{\alpha}}^{\frac{2}{\alpha},0}\left[\begin{array}{c}- \\
\mathbf{b}
\end{array}\biggr|\; x^{2}\right]\label{eq:surface-mass-einasto-m-interger}
\end{equation}

\noindent 
where $\mathbf{b}$ is a vector of size $\frac{2}{\alpha}$ given by:

\begin{equation}
\mathbf{b}=\biggl\{\frac{\alpha}{2},2\frac{\alpha}{2},...,
\left(\frac{2}{\alpha}-1\right)\frac{\alpha}{2},-\frac{1}{2}\biggr\}
\end{equation}

\noindent 
This result indicates that the form of the surface mass
density of the Einasto profile differs from the surface mass density
of the S\'ersic model (equation \ref{eq:Sersic-surface-mass-density})
in functional form.

\subsubsection{Einasto index with rational values}

\noindent 
Also is possible to write this expressions for all rational
values of the Einasto index. Using $m=p/q$ with $p$ and $q$ both
integer numbers equation (\ref{eq:Einasto-mellin-transform}) becomes:

\begin{equation}
\Sigma\left(x\right)=\rho_{-2}r_{-2}\sqrt{\pi}e^{2m}\: x\:\frac{1}{2\pi i}
\int_{_{\mathit{C}}}\frac{q\Gamma\left(-\frac{1}{2}+qy\right)
\Gamma\left(1+2py\right)}{\Gamma\left(1+qy\right)}\left[
\left(\frac{2p}{q}\right)^{2p}x^{2q}\right]^{-y}dy 
\label{eq:Mellin-transform-m-rational}
\end{equation}

\noindent 
Substituting the three Gamma functions in equation 
(\ref{eq:Mellin-transform-m-rational}) using the equation 
(\ref{eq:primer-gamma}), we obtain an integral and compare it with the 
definition of the Meijer G-function, we find:

\begin{equation}
\Sigma\left(x\right)=\frac{\rho_{-2}r_{-2}e^{\frac{2p}{q}}}{
\left(2\pi\right)^{p-1}}\sqrt{\frac{p}{q}}\: x\: 
G_{q-1,2p+q-1}^{2p+q-1,0}\left[\begin{array}{c}
\mathbf{a}\\
\mathbf{b}
\end{array}\biggr|\;\frac{x^{2q}}{q^{2p}}\right] 
\label{eq:surface-mass-einasto-rational}
\end{equation}

\noindent 
where $\mathbf{a}$ and $\mathbf{b}$ are vectors of size
$q-1$ and $2p+q-1$ respectively given by:

\begin{equation}
\mathbf{a}=\biggl\{\frac{1}{q},\frac{2}{q},...,\frac{q-1}{q}\biggr\}
\end{equation}

\begin{equation}
\mathbf{b}=\biggl\{\frac{1}{2p},\frac{2}{2p},...,\frac{2p-1}{2},-\frac{1}{2q},
\frac{1}{2q},\frac{3}{2q},...,\frac{2q-3}{2q}\biggr\}
\end{equation}

\noindent 
It is immediate to verify that the equation 
~(\ref{eq:surface-mass-einasto-rational}) is equivalent to the equation 
~(\ref{eq:surface-mass-einasto-m-interger}) for Einasto index with values 
$\alpha_{E}=\frac{1}{n}$. Using the properties of the Meijer G-function 
~(\ref{eq:Meijer-reduction-011}, \ref{eq:Meijer-reduction-012}) 
and (\ref{eq:Meijer-reduction-02})
is possible to demonstrate that equations 
~(\ref{eq:surface-mass-einasto-rational}) and 
~(\ref{eq:surface-mass-einasto-m-interger}) are equal for Einasto index with 
values $\alpha_{E}=\frac{2}{n}$.

\subsubsection{Simple cases: $\alpha_{E}=1$ and $\alpha_{E}=2$ }

\noindent 
In order to check our results, we projected the Einasto profile in
the cases $\alpha=1$ and $\alpha=2$ and compared the results with
equation 
~(\ref{eq:surface-mass-einasto-m-interger}). 
In both cases the values of $\alpha$
are outside the range favored by the N-body CDM simulations, but are
practical to check the consistency of our calculations. 

\noindent 
For the case $\alpha=1$ we have:

\begin{equation}
\rho\left(r\right)=\rho_{-2}\exp\left(-2\left[\frac{r}{r_{-2}}-1\right]\right)
\end{equation}

\noindent 
Calculating the projected surface mass density using the equation 
(\ref{eq:04}), we find:

\begin{equation}
\Sigma\left(x\right)=2\rho_{-2}r_{-2}e^{2}\: x\: K_{1}\left(2x\right)
\label{eq:surface-mass-bessel-einasto-index=00003D1}
\end{equation}

\noindent 
where $K_{1}\left(x\right)$ is the modified Bessel of the second kind.

\noindent 
Setting $\alpha=1$ in equation ~(\ref{eq:surface-mass-einasto-m-interger}), 
we have:

\begin{equation}
\Sigma\left(x\right)=\rho_{-2}r_{-2}e^{2}\: x\: G_{0,2}^{2,0}
\left[\begin{array}{c}-\\ 
\frac{1}{2},-\frac{1}{2}
\end{array}\biggr|\; x^{2}\right]
\label{eq:surface-mass-meijer-einasto-index=00003D1}
\end{equation}

\noindent 
Substituting the equation (\ref{eq:bessel_second-meijer})
into (\ref{eq:surface-mass-meijer-einasto-index=00003D1}) we obtain
the equation (\ref{eq:surface-mass-bessel-einasto-index=00003D1}).

\noindent 
In a similar way with the case $\alpha=2$:

\begin{equation}
\rho\left(r\right)=\rho_{-2}\exp\left(-\left[\left(
\frac{r}{r_{-2}}\right)^{2}-1\right]\right)
\end{equation}

\noindent 
The projected surface mass density can be found using the equation 
(\ref{eq:04}):

\begin{equation}
\Sigma\left(x\right)=\sqrt{\pi}\rho_{-2}r_{-2}e^{-x^{2}+1}
\label{eq:surface-mass-exponential-einasto-index=00003D2}
\end{equation}

\noindent 
Setting $\alpha=2$ in equation ~(\ref{eq:surface-mass-einasto-m-interger}), 
we have:

\begin{equation}
\Sigma\left(x\right)=\sqrt{\pi}\rho_{-2}r_{-2}e^{1}\: x\: G_{0,1}^{1,0}
\left[\begin{array}{c}-\\
-\frac{1}{2}
\end{array}\biggr|\; x^{2}\right]
\label{eq:surface-mass-meijer-einasto-index=00003D2}
\end{equation}

\noindent 
Using (\ref{eq:exponetial-meijer}) in the last equation this one reduces to 
equation (\ref{eq:surface-mass-exponential-einasto-index=00003D2}).

\begin{figure}
\begin{centering}
\includegraphics[scale=0.8]{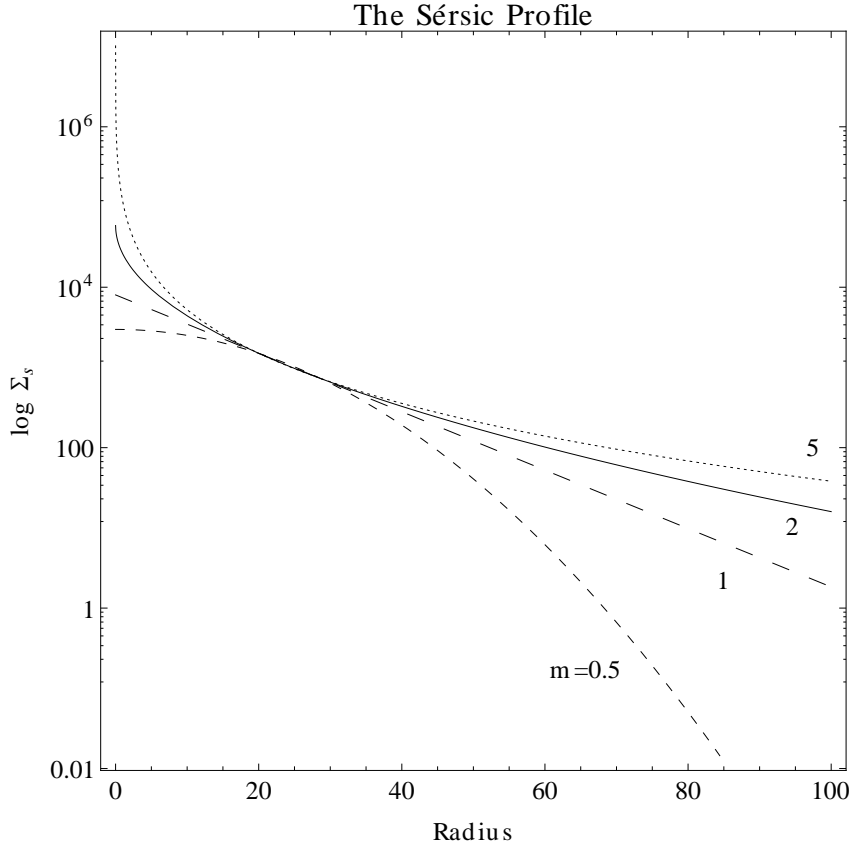}
\caption{S\'ersic profile where $\Upsilon I_{e}$ and $R_{e}$ are held fixed 
for four values of the S\'ersic index $m$.
\label{fig:plot01-Sersic-profile}}
\end{centering}
\end{figure}

\begin{figure}
\begin{centering}
\includegraphics[scale=0.8]{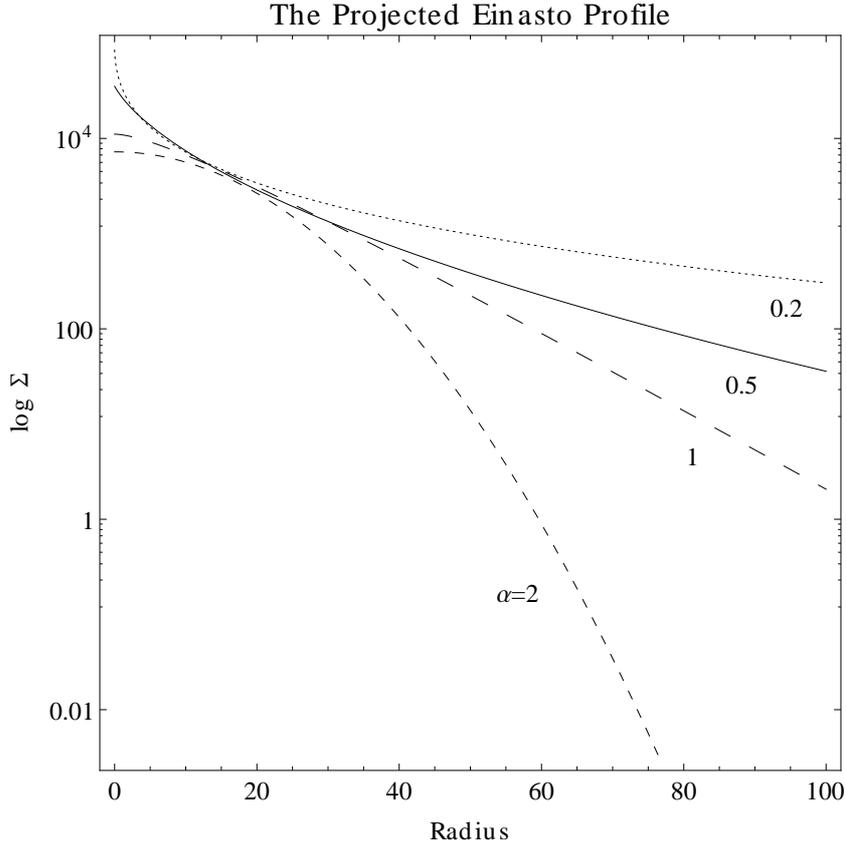}
\caption{Projected Einasto profile where $\rho_{-2}r_{-2}$ and $r_{-2}$ are held fixed
for four values of the Einasto index $\alpha$.
\label{fig:plot02-Einasto-profile}}
\end{centering}
\end{figure}

\noindent 
It is interesting to compare these two cases with the surface mass
density of the S\'ersic profile $\Sigma_{S}\left(R\right)$ with the
same values for the S\'ersic index $1/m$ that for the Einasto index
$\alpha$. We also include the cases $\alpha=0.5\:(m=2)$ and $\alpha=0.2\:(5)$
for the comparison. The Figure \ref{fig:plot01-Sersic-profile} shows
$\Sigma_{S}\left(R\right)$ for four values of $m$ and Figure 
\ref{fig:plot02-Einasto-profile} displays $\Sigma\left(x\right)$ for four 
values of $\alpha$. In both it can be seen clearly that the respective index 
is very important in determining the overall behavior of the curves. 
The S\'ersic profile is characterized by a more steeper central core and 
extended external wing for larger values of the S\'ersic index $m$. 
For low values of $m$ the central core is more flat and the external wing is 
sharply truncated. The Einasto profile has a similar behavior, with the 
difference that the external wings are most spread out. Also in the inner 
region for both profiles with low values of the respectively index we obtain
larger values of $\Sigma_{S}$ and $\Sigma$. However, the Einasto
profile seems to be less sensitive to the value of the surface mass
density for a given $\alpha$ and radius and in the inner region than
the S\'ersic profile. It is in this region where the lensing effect
is more important and the difference in the surface mass density determines
the lensing properties of the respectively profiles. Given this difference,
we see that the lensing properties of the S\'ersic and Einasto profile
are not equal. Studies of the lensing properties of the S\'ersic profile
had been done by \citet{2004A&A...415..839C} and \citet{2007JCAP...07..006E}.

\section{The total mass enclosed \label{sec:03}}

\noindent 
The total mass enclosed in a halo described by the Einasto profile
can be found by:

\begin{equation}
M_{tot}=4\pi\int_{0}^{\infty}\rho\left(r\right)r^{2}dr
\label{eq:_integral_total_mass_spherical}
\end{equation}

\noindent 
Combining equations (\ref{eq:02}) and 
(\ref{eq:_integral_total_mass_spherical}), we get:

\begin{equation}
M_{tot}=\frac{4\pi\rho_{-2}r_{-2}^{3}e^{\frac{4}{\alpha}}}{\alpha}\left(
\frac{\alpha}{2}\right)^{\frac{3}{\alpha}}\Gamma\left(\frac{3}{\alpha}\right)
\label{eq:total_mass_einasto}
\end{equation}

\noindent 
This result was also obtained by \citet{2005MNRAS.358.1325C}. 

\noindent 
We can get the same result calculating the total mass projected
on the sky plane:

\begin{equation}
M_{tot}=2\pi\int_{0}^{\infty}\Sigma\left(\xi\right)\xi d \xi\label{eq:total_mass_2D}
\end{equation}

\noindent 
Inserting equation ~(\ref{eq:surface-mass-einasto-rational}) into 
(\ref{eq:total_mass_2D}), we find:

\begin{equation}
M_{tot}=\frac{\rho_{-2}r_{-2}^{3}e^{\frac{2p}{q}}}{2q\left(2\pi\right)^{p-2}}
\sqrt{\frac{p}{q}}\int_{0}^{\infty}G_{q-1,2p+q-1}^{2p+q-1,0}
\left[\begin{array}{c}
\mathbf{a}\\
\mathbf{b}
\end{array}\biggr|\;\frac{x^{\prime}}{q^{2p}}\right]\left(
x^{\prime}\right)^{\frac{3}{2q}-1}dx^{\prime}
\end{equation}

\noindent 
Integrating the last equation using the formula 
(\ref{eq:indefinite-integration-meijer}) for indefinite integration of the 
Meijer G-function: 

\begin{equation}
M_{tot}=\frac{\rho_{-2}r_{-2}^{3}e^{\frac{2p}{q}}}{2q\left(2\pi\right)^{p-2}}
\sqrt{\frac{p}{q}}\left(q^{2p}\right)^{\frac{3}{2q}}\frac{\prod_{j=1}^{2p+q-1}
\Gamma\left(\frac{3}{2q}+b_{j}\right)}{\prod_{j=1}^{q-1}
\Gamma\left(\frac{3}{2q}+a_{j}\right)} \label{eq:total_mass_products}
\end{equation}

\noindent 
We can write both products appearing in the numerator and
denominator in equation (\ref{eq:total_mass_products}) using the
Gauss multiplication formula (\ref{eq:gauss_mult_form}) respectively
as:

\begin{eqnarray}
& \prod_{j=1}^{2p+q-1} & \Gamma\left(\frac{3}{2q}+b_{j}\right)=
\frac{\prod_{j=0}^{2p-1}\Gamma\left(\frac{\frac{3}{2q}+j}{2p}\right)
\prod_{j=0}^{q-1}\Gamma\left(\frac{1+j}{q}\right)}{
\Gamma\left(\frac{3}{2q}\right)} \nonumber \\
& = & 
\sqrt{\pi}\left(2\pi\right)^{p+\frac{q}{2}-1}\left(2p\right)^{\frac{1}{2}
-\frac{3p}{q}}\frac{\Gamma\left(\frac{3p}{q}\right)}{q^{\frac{1}{2}}
\Gamma\left(\frac{3}{2q}\right)}
\label{eq:product_01}
\end{eqnarray}

\begin{equation}
\prod_{j=1}^{q-1}\Gamma\left(\frac{3}{2q}+a_{j}\right)=\frac{\prod_{j=0}^{q-1}
\Gamma\left(\frac{\frac{3}{2}+j}{q}\right)}{\Gamma\left(\frac{3}{2q}\right)}=
\frac{\sqrt{\pi}q^{-1}\left(2\pi\right)^{\frac{q-1}{2}}}{
2\Gamma\left(\frac{3}{2q}\right)} \label{eq:product_02}
\end{equation}

\noindent 
Substituting equations (\ref{eq:product_01}) and (\ref{eq:product_02})
into (\ref{eq:total_mass_products}) with $\alpha=\frac{q}{p}$ we
obtain the same result (\ref{eq:total_mass_einasto}) for the total
mass enclosed of the Einasto profile. This confirms that our calculations
for the surface mass density of the Einasto profile are correct.

\section{The deflection angle, lens equation and the lensing potential
\label{sec:04}}

\subsection{The deflection angle and lens equation:}

\noindent 
In the thin lens approximation, the lens equation for a axially symmetric
lens is:

\begin{equation}
\eta=\frac{D_{S}}{D_{L}}\xi-D_{LS}\hat{\alpha}
\label{eq:lens-equation-dimensional}
\end{equation}

\noindent 
where the quantities $\eta$ and $\xi$ are the physical positions
of the  of a source in the source plane and an image in the image
plane, respectively, $\hat{\alpha}$ is the deflection angle, and
$D_{L}$, $D_{S}$ and $D_{LS}$ are the angular distances from observer
to lens, observer to source, and lens to source, respectively.

\noindent 
With the dimensionless positions $y=D_{L}\eta/D_{S}r_{-2}$
and $x=\xi/r_{-2}$, and dimensionless $\alpha=D_{L}D_{LS}\hat{\alpha}/D_{S}\xi$
the lens equation reduces to:

\begin{equation}
y=x-\alpha\left(x\right)\label{eq:lens-equation-adimensional}
\end{equation}

\noindent 
The deflection angle for a spherical symmetric lens is 
\citep{1992grle.book.....S}:

\begin{equation}
\alpha\left(x\right)=\frac{2}{x}\int_{0}^{x}x^{\prime}
\frac{\Sigma\left(x^{\prime}\right)}{\Sigma_{crit}}dx^{\prime}=
\frac{2}{x}\int_{0}^{x}x^{\prime}\kappa\left(x^{\prime}\right)dx^{\prime}
\label{eq:deflection-angle}
\end{equation}

\noindent 
where $\kappa\left(x\right)=\Sigma\left(x\right)/\Sigma_{crit}$
is the convergence and $\Sigma_{crit}$ is the critical surface mass
density defined by:

\begin{equation}
\Sigma_{crit}=\frac{c^{2}D_{S}}{4\pi GD_{L}D_{LS}}
\end{equation}

\noindent 
where $c$ is the speed of light, $G$ is the gravitational constant.

\noindent 
Inserting equation ~(\ref{eq:surface-mass-einasto-rational}) into 
(\ref{eq:deflection-angle}), we find the deflection angle of the Einasto 
profile:

\begin{equation}
\alpha\left(x\right)=\frac{\rho_{-2}r_{-2}e^{\frac{2p}{q}}}{
\left(2\pi\right)^{p-1}q\Sigma_{crit}}\sqrt{\frac{p}{q}}\: x^{2}\: 
G_{q,2p+q}^{2p+q-1,1}\left[\begin{array}{c} 1-\frac{3}{2q},\mathbf{a}\\
\mathbf{b},-\frac{3}{2q}
\end{array}\biggr|\;\frac{x^{2q}}{q^{2p}}\right]
\end{equation}

\noindent 
Introducing the central convergence, $\kappa_{c}$, a parameter
that determinate the lensing properties of the Einasto profile, defined by:

\begin{equation}
\kappa_{c}\equiv\frac{\Sigma\left(x=0\right)}{\Sigma_{crit}}=
\frac{\rho_{-2}r_{-2}e^{\frac{2p}{q}}\Gamma\left(\frac{p}{q}\right)}{
\Sigma_{crit}}\left(\frac{q}{2p}\right)^{\frac{p}{q}-1}=\frac{\rho_{-2}r_{-2}
e^{\frac{2}{\alpha}}\Gamma\left(\frac{1}{\alpha}\right)}{\Sigma_{crit}}
\left(\frac{\alpha}{2}\right)^{\frac{1}{\alpha}-1}
\end{equation}

\noindent 
and use it to write $\alpha\left(x\right)$ in terms of $\kappa_{c}$:

\begin{equation}
\alpha\left(x\right)=\frac{\kappa_{c}}{\left(2\pi\right)^{p-1}
\Gamma\left(\frac{p}{q}\right)q}\sqrt{\frac{p}{q}}
\left(\frac{2p}{q}\right)^{\frac{p}{q}-1}\: x^{2}\: 
G_{q,2p+q}^{2p+q-1,1}\left[\begin{array}{c}
1-\frac{3}{2q},\mathbf{a}\\
\mathbf{b},-\frac{3}{2q}
\end{array}\biggr|\;\frac{x^{2q}}{q^{2p}}\right]
\label{eq:deflection-angle-meijer-rational}
\end{equation}

\noindent 
For Einasto index with values $\alpha=\frac{1}{n}$ and $\alpha=\frac{2}{n}$
with $n$ integer, the last equation can be written as:

\begin{equation}
\alpha\left(x\right)=\frac{\kappa_{c}}{\left(2\pi\right)^{\frac{1}{\alpha}-1}
\Gamma\left(\frac{1}{\alpha}\right)\sqrt{\alpha}}
\left(\frac{2}{\alpha}\right)^{\frac{1}{\alpha}-1}\: x^{2}\: 
G_{1,\frac{2}{\alpha}+1}^{\frac{2}{\alpha},1}\left[\begin{array}{c}
-\frac{1}{2},\mathbf{a}\\
\mathbf{b},-\frac{3}{2}
\end{array}\biggr|\; x^{2}\right]
\end{equation}

\noindent 
The lens equation for the Einasto profile is then:

\begin{equation}
y=x-\frac{\kappa_{c}}{\left(2\pi\right)^{p-1}\Gamma\left(\frac{p}{q}\right)q}
\sqrt{\frac{p}{q}}\left(\frac{2p}{q}\right)^{\frac{p}{q}-1}\: x^{2}\: 
G_{q,2p+q}^{2p+q-1,1}\left[\begin{array}{c}
1-\frac{3}{2q},\mathbf{a}\\
\mathbf{b},-\frac{3}{2q}
\end{array}\biggr|\;\frac{x^{2q}}{q^{2p}}\right]
\end{equation}

\noindent 
which can be simplified to:

\begin{equation}
y=x-\frac{\kappa_{c}}{\left(2\pi\right)^{\frac{1}{\alpha}-1}
\Gamma\left(\frac{1}{\alpha}\right)\sqrt{\alpha}}
\left(\frac{2}{\alpha}\right)^{\frac{1}{\alpha}-1}\: x^{2}\: 
G_{1,\frac{2}{\alpha}+1}^{\frac{2}{\alpha},1}\left[\begin{array}{c}
-\frac{1}{2},\mathbf{a}\\
\mathbf{b},-\frac{3}{2}
\end{array}\biggr|\; x^{2}\right]
\end{equation}

\noindent 
for Einasto index with values $\alpha=\frac{1}{n}$ and
$\alpha=\frac{2}{n}$ with $n$ integer.

\noindent 
For a spherically symmetric lens being capable of forming multiple
images of the source a sufficient condition is $\kappa_{c}>1$ 
\citep{1992grle.book.....S}.
In the case $\kappa_{c}\leq1$ only one image of the source is formed.
In addition to the condition $\kappa_{c}>1$ multiples images are
produced only if $\mid y\mid\leq y_{crit}$ \citep{2002ApJ...566..652L},
where $y_{crit}$ is the the maximum value of $y$ when $x<0$ or
the minimum for $x>0$. For singular profiles such as the NFW profile,
the central convergence always is divergent, hence the condition $\kappa_{c}>1$
is always met, this implies that the NFW profile is capable of forming
multiple images for any mass. Nonsingular profiles such as the Einasto
profile are not capable of forming multiple images for any mass. Instead,
the condition $\kappa_{c}>1$ sets a threshold for the lens mass required
to form multiple images.

\subsection{The deflection potential}

\noindent 
The deflection potential $\psi\left(x\right)$ for spherically symmetric
lens is given by:

\begin{equation}
\alpha\left(x\right)=\frac{d\psi}{dx}\label{eq:deflection_angle_potential}
\end{equation}

\noindent 
We see from equation (\ref{eq:deflection_angle_potential}) that can
find the lensing potential simply integrating the deflection angle:

\begin{equation}
\psi\left(x\right)=\int_{0}^{x}\alpha\left(x^{\prime}\right)dx^{\prime}
\label{eq:potential_deflection_angle}
\end{equation}

\noindent 
Inserting the equation (\ref{eq:deflection-angle-meijer-rational})
into (\ref{eq:potential_deflection_angle}) and using the identity
(\ref{eq:indefinite-integration-meijer}), we found:

\begin{equation}
\psi\left(x\right)=\frac{\kappa_{c}}{2\left(2\pi\right)^{p-1}
\Gamma\left(\frac{p}{q}\right)q^{2}}\sqrt{\frac{p}{q}}
\left(\frac{2p}{q}\right)^{\frac{p}{q}-1}\: x^{3}\: 
G_{q+1,2p+q+1}^{2p+q-1,2}\left[\begin{array}{c}
1-\frac{3}{2q},1-\frac{3}{2q},\mathbf{a}\\
\mathbf{b},-\frac{3}{2q},-\frac{3}{2q}
\end{array}\biggr|\;\frac{x^{2q}}{q^{2p}}\right]
\end{equation}

\noindent 
which can be reduced to :

\begin{equation}
\psi\left(x\right)=\frac{\kappa_{c}}{2\left(2\pi\right)^{\frac{1}{\alpha}-1}
\Gamma\left(\frac{1}{\alpha}\right)\sqrt{\alpha}}
\left(\frac{2}{\alpha}\right)^{\frac{1}{\alpha}-1}\: x^{3}\: 
G_{2,\frac{2}{\alpha}+2}^{\frac{2}{\alpha},2}\left[\begin{array}{c}
-\frac{1}{2},-\frac{1}{2},\mathbf{a}\\
\mathbf{b},-\frac{3}{2},-\frac{3}{2}
\end{array}\biggr|\; x^{2}\right]
\end{equation}

\noindent 
for Einasto index with values $\alpha=\frac{1}{n}$ and
$\alpha=\frac{2}{n}$ with $n$ integer.

\section{Magnification, shear and the critical curves \label{sec:05}}

\noindent 
The gravitational lensing effect preservers the surface brightness
but causes variations in the shape and the solid angle of the source.
Thereby, the source luminosity is amplified by \citep{1992grle.book.....S}:

\begin{equation}
\mu=\frac{1}{\left(1-\kappa\right)^{2}-\gamma^{2}}\label{eq:magnification}
\end{equation}

\noindent 
where $\kappa\left(x\right)$
is the convergence and $\gamma\left(x\right)$ is the shear. The amplification
has two contributions one from the convergence which describes an
isotropic focusing of light rays in the lens plane and the other is
an anisotropic focusing caused by the tidal gravitational forces acting
on the light rays, described by the shear. For a spherical symmetric
lens, the shear is given by \citep{1991ApJ...370....1M}:

\begin{equation}
\gamma\left(x\right)=\frac{\bar{\Sigma}\left(x\right)
-\Sigma\left(x\right)}{\Sigma_{crit}}=\bar{\kappa}-\kappa\label{eq:shear}
\end{equation}

\noindent 
where

\begin{equation}
\bar{\Sigma}\left(x\right)=\frac{2}{x^{2}}\int_{0}^{x}x^{\prime}
\Sigma\left(x^{\prime}\right)dx^{\prime} \label{eq:mean-surface-mass-density}
\end{equation}

\noindent 
is the average surface mass density within $x$.

\noindent 
The magnification of the Einasto profile can be found combining equations
~(\ref{eq:surface-mass-einasto-rational}), (\ref{eq:magnification}), 
(\ref{eq:shear}) and (\ref{eq:mean-surface-mass-density}). We get:

\begin{equation}
\mu=\left[\left(1-\bar{\kappa}\right)\left(1+\bar{\kappa}
-2\kappa\right)\right]^{-1} \label{eq:magnification-einasto}
\end{equation}

\noindent 
where 

\begin{equation}
\kappa\left(x\right)=\frac{\kappa_{c}}{\left(2\pi\right)^{p-1}
\Gamma\left(\frac{p}{q}\right)}\sqrt{\frac{p}{q}}
\left(\frac{2p}{q}\right)^{\frac{p}{q}-1}\: x\: 
G_{q-1,2p+q-1}^{2p+q-1,0}\left[\begin{array}{c}
\mathbf{a}\\
\mathbf{b}
\end{array}\biggr|\;\frac{x^{2q}}{q^{2p}}\right]
\end{equation}

\begin{equation}
\bar{\kappa}\left(x\right)=\frac{\kappa_{c}}{\left(2\pi\right)^{p-1}
\Gamma\left(\frac{p}{q}\right)q}\sqrt{\frac{p}{q}}
\left(\frac{2p}{q}\right)^{\frac{p}{q}-1}\: x\: 
G_{q,2p+q}^{2p+q-1,1}\left[\begin{array}{c}
1-\frac{3}{2q},\mathbf{a}\\
\mathbf{b},-\frac{3}{2q}
\end{array}\biggr|\;\frac{x^{2q}}{q^{2p}}\right]
\end{equation}

\noindent
The last equations reduce to:

\begin{equation}
\kappa\left(x\right)=\frac{\kappa_{c}}{\left(2\pi\right)^{\frac{1}{\alpha}-1}
\Gamma\left(\frac{1}{\alpha}\right)\sqrt{\alpha}}
\left(\frac{2}{\alpha}\right)^{\frac{1}{\alpha}-1}\: x\: 
G_{0,\frac{2}{\alpha}}^{\frac{2}{\alpha},0}\left[\begin{array}{c}
\mathbf{a}\\
\mathbf{b}
\end{array}\biggr|\; x^{2}\right]
\end{equation}

\begin{equation}
\bar{\kappa}\left(x\right)=\frac{\kappa_{c}}{
\left(2\pi\right)^{\frac{1}{\alpha}-1}\Gamma\left(\frac{1}{\alpha}\right)
\sqrt{\alpha}}\left(\frac{2}{\alpha}\right)^{\frac{1}{\alpha}-1}\: x\: 
G_{1,\frac{2}{\alpha}+1}^{\frac{2}{\alpha},1}\left[\begin{array}{c}
-\frac{1}{2},\mathbf{a}\\
\mathbf{b},-\frac{3}{2}
\end{array}\biggr|\; x^{2}\right]
\end{equation}

\noindent 
for Einasto index with values $\alpha=\frac{1}{n}$ and
$\alpha=\frac{2}{n}$ with $n$ integer.

\noindent 
The magnification may be divergent for some image positions. The loci
of the divergent magnification in the image plane are called the critical
curves. For the Einasto profile we see from equation 
(\ref{eq:magnification-einasto})
that has one pair of critical curves. The first curve $1-\bar{\kappa}=0$
is the tangential critical curve which correspond to an Einstein Ring
with a radius called the Einstein radius. The second curve 
$1+\bar{\kappa}-2\kappa=0$
is the radial critical curve which also defines a ring and its correspond
radius. In both cases the equations must be solved numerically.

\section{Summary and Conclusions \label{sec:06}}

\noindent 
In this paper, we have derived an analytical expression for the surface
mass density of the Einasto profile using the Mellin transformed.
This expression can be written in terms of the Fox H-function for
general values of the Einasto index $\alpha$. The same expression
can be written in terms of the Meijer G-function for all rational
values of the Einasto index, with a simplification for values 
$\alpha=\frac{1}{n}$ and $\alpha=\frac{2}{n}$ with $n$ integer of the Einasto 
index. One we obtained an analytical expression for the surface mass density
we also derived in terms of the Meijer G-function other lensing properties:
deflection angle, lens equation, deflection potential, magnification,
shear and critical curves of the Einasto profile for all rational
values of the Einasto index, with a simplification for values 
$\alpha=\frac{1}{n}$ and $\alpha=\frac{2}{n}$ with $n$ integer of the Einasto 
index. Our analytical results can be used to investigate further the lensing
properties of the Einasto profile taking advantage of the fact that
the Meijer G-function is a very well studied function in the literature.

\noindent 
We compared the S\'ersic and Einasto surface mass density profiles using
the equivalent values for the S\'ersic $m$ and Einasto $\alpha$ indexes
and where the quantities $\Upsilon I_{e}$, $R_{e}$ and $\rho_{-2}r_{-2}$,
$r_{-2}$ are held fixed. We found that both profiles have similar
behavior determined by the index value. However, we noted that for
the Einasto profile the external wings are most spread out and seems
to be less sensitive to the value of the surface mass density for
a given Einasto index and radius in the inner region than the S\'ersic
profile. This feature is key because it is in this region where the
lensing effect is more important and the difference of the surface
mass densities implies a difference in the lensing properties of the
two profiles.

\noindent 
Our results can be used in strong and weak lensing studies of galaxies
and clusters where dark matter is to believed the main mass component
and the mass distribution can be assumed to be given by the Einasto
profile. The implementation of this results is easy because the Meijer
G-function is available in several commercial and open-source CAS.
The performance of this nonsingular three-parameter model in fitting
the 3D spatial densities in high resolution N-body CDM simulations
is better than the singular two-parameter NFW profile makes very promising
its use in strong and weak lensing studies. The constant increasing
computational power available opens the possibility of using most
realistic and sophisticated profiles like the Einasto profile for
lensing studies and marks a route to obtain a satisfactory solution
to the cusp-core problem.

\bigskip
\noindent
{\bf Acknowledgements}: The authors wish to thank H. Morales and R. Carboni 
for critical reading. This research has made use of NASA's Astrophysics Data 
System Bibliographic Services.  

\bibliographystyle{aa}
\bibliography{my_bib}

\appendix
\section{The Meijer transform-method}  \label{sec:A}

\noindent 
The Mellin transform-method \citep{0853125287,1996MER.16..05,159829184X}
uses the Mellin integral transform for the integral evaluation.

\noindent 
The Mellin transform of a function $f\left(z\right)$ is
an integral transform defined by:

\begin{equation}
\left\{ \mathcal{M}f\right\} \left(u\right)=
\int_{0}^{\infty}z^{u-1}f\left(z\right)dz
\end{equation}

\noindent 
if the integral exits.

\noindent 
It is clear from the definition that the Mellin transform
does not exist for all functions such as the polynomials, the integral
does not converge. The Mellin transform when it does exits it converges
in a vertical strip in the complex {$z$}-plane. This strip is
called the {\it strip of analyticity} (SOA).

\noindent 
The inverse Mellin transform is defined by: 

\begin{equation}
f\left(z\right)=\frac{1}{2\pi i}\int_{_{\mathit{C}}}z^{-u}\left\{ 
\mathcal{M}f\right\} \left(u\right)du
\end{equation}

\noindent 
where the contour of integration $C$ is a vertical line
in the complex {$z$}-plane and must be placed in the SOA of
$f\left(z\right)$. 

\noindent 
Given two functions $f\left(z\right)$ and $g\left(z\right)$
the Mellin convolution is defined by:

\begin{equation}
\left(f\star g\right)\left(z\right)=\int_{0}^{\infty}f\left(y\right)
g\left(\frac{z}{y}\right)\frac{dy}{y}
\end{equation}

\noindent 
It is well know that the Laplace or Fourier transform of
the product of two different functions is the convolution of the respectively
transform. In the case of the Mellin transform we have:

\begin{equation}
\int_{0}^{\infty}f\left(y\right)g\left(\frac{z}{y}\right)\frac{dy}{y}=
\frac{1}{2\pi i}\int_{_{\mathit{C}}}z^{-u}\left\{ \mathcal{M}f\right\} 
\left(u\right)\left\{ \mathcal{M}g\right\} \left(u\right)du
\label{eq:central_feature_mellin}
\end{equation}

\noindent 
if $z=1$ this formula is know as the Parseval's theorem
for the Mellin transform.

\noindent 
The most important feature of the Mellin transform-method
is that using the equation (\ref{eq:central_feature_mellin}) integrals
of the type,

\begin{equation}
I\left(z\right)=\int_{0}^{\infty}f\left(y\right)g\left(\frac{z}{y}\right)
\frac{dy}{y}
\end{equation}

\noindent 
can be written as an inverse Mellin transform. With the
requirement that $f$ and $g$ should be of the hypergeometric type
and consequently their Mellin transforms can be written as products
with the form $\Gamma\left(a+Au\right)$ or 
$\left[\Gamma\left(a+Au\right)\right]^{-1}$
with the $A$'s being real numbers, the resulting integrals are of
the Mellin-Barnes type and then can be written in terms of the Fox
H-function for $A\neq1$ or the Meijer G-function for $A=1$ 
(see Appendix \ref{sec:B}).

\section{The Meijer G function and its properties}  \label{sec:B}

\noindent 
The Meijer G-function is a very general, analytical
function introduced by \citet{1936NAvW.18..10} which includes most
of the special functions as specific cases. It is defined in terms
of the inverse Mellin transform \citep{9780070195493} by: 

\begin{eqnarray}
& G_{p,q}^{m,n} &
\left[\begin{array}{c}
\mathbf{a}\\
\mathbf{b}
\end{array}\biggr|\; z\right]\equiv G_{p,q}^{m,n}\left[\begin{array}{c}
a_{1},...,a_{p}\\
b_{1},...,b_{q}
\end{array}\biggr|\; z\right] \nonumber \\ 
& = & 
\frac{1}{2\pi i}
\int_{_{\mathit{C}}}\frac{{\textstyle 
\prod_{j=1}^{m}\Gamma\left(b_{j}+s\right)\prod_{j=1}^{n}
\Gamma\left(1-a_{j}-s\right)}}{{\textstyle \prod_{j=m+1}^{q}
\Gamma\left(1-b_{j}-s\right)\prod_{j=n+1}^{p}
\Gamma\left(a_{j}+s\right)}}z^{-s}ds \label{eq:Meijer-G}
\end{eqnarray}

\noindent 
where $\mathit{C}$ is a contour in the complex plane, $\Gamma(s)$
is the Gamma function and $\mathbf{a}$ and $\mathbf{b}$ are vectors
of dimension $p$ and $q$, respectively.

\noindent 
The basic properties of the Meijer G-function are
too numerous to be mention here. We only provide a short list of the
most relevant properties for this work.

\noindent 
A Meijer G-function with $p>q$ can be transformed 
to another G-function with $p<q$ :

\begin{equation}
G_{p,q}^{m,n}\left[\begin{array}{c}
\mathbf{a}\\
\mathbf{b}
\end{array}\biggr|\; z\right]=G_{q,p}^{m,n}\left[\begin{array}{c}
1-b_{1},...,1-b_{q}\\
1-a_{1},...,1-a_{p}
\end{array}\biggr|\;\frac{1}{z}\right]
\end{equation}

\noindent 
Other property is that if one the parameters of $\mathbf{a}$
and $\mathbf{b}$ appears in both the numerator and denominator of
the integrand, the order of the Meijer G-function may decrease
and the fraction simplified. The positions of the parameters dictates
which order $m$ or $n$ will decrease. For example if $a_{k}=b_{j}$
for some $k=1,2,...,n$ and $j=m+1,m+2,...,q$ , the orders $p,q$
and $n$ of the Meijer G-function will decrease:

\begin{equation}
G_{p,q}^{m,n}\left[\begin{array}{c}
a_{1},a_{2},...,a_{p}\\
b_{1},...b_{q-1},a_{1}
\end{array}\biggr|\; z\right]=G_{p-1,q-1}^{m,n-1}\left[\begin{array}{c}
a_{2},...,a_{p}\\
b_{1},...,b_{q-1}
\end{array}\biggr|\; z\right]
\label{eq:Meijer-reduction-011}
\end{equation}

\noindent 
In the other case if $a_{k}=b_{j}$ for some $k=n+1,n+2,...,p$
and $j=1,2,...,m$, the orders $p,q$ and $m$ of the Meijer G-function will 
decrease:

\begin{equation}
G_{p,q}^{m,n}\left[\begin{array}{c}
a_{1},...,a_{p-1},b_{1}\\
b_{1},b_{2},...b_{q}
\end{array}\biggr|\; z\right]=G_{p-1,q-1}^{m-1,n}\left[\begin{array}{c}
a_{1},...,a_{p-1}\\
b_{2},...,b_{q}
\end{array}\biggr|\; z\right]
\label{eq:Meijer-reduction-012}
\end{equation}

\noindent 
The order reduction formula for the Meijer G-function is:

\begin{eqnarray}
& G_{p,q}^{m,n} &
\left[\begin{array}{c}
a_{1},...,a_{p}\\
b_{1},...,b_{q}
\end{array}\biggr|\; z\right]=
\frac{k^{1+v+\left(p-q\right)/2}}{\left(2\pi\right)^{\left(k-1\right)\delta}} 
\\
& \times & G_{kp,kq}^{km,kn}\left[\begin{array}{c}
a_{1}/k,...,\left(a_{1}+k-1\right)/k,...,a_{p}/k,...,\left(a_{p}+k-1\right)/k\\
b_{1}/k,...,\left(b_{1}+k-1\right)/k,...,b_{q}/k,...,\left(b_{q}+k-1\right)/k
\end{array}\biggr|\;\frac{z^{k}}{k^{k\left(p-q\right)}}\right]
\nonumber 
\label{eq:Meijer-reduction-02}
\end{eqnarray}

\noindent 
The multiplication by powers of $z$ is another property:

\begin{equation}
z^{\alpha}G_{p,q}^{m,n}\left[\begin{array}{c}
a_{1},...,a_{p}\\
b_{1},...,b_{q}
\end{array}\biggr|\; z\right]=G_{p,q}^{m,n}\left[\begin{array}{c}
a_{1}+\alpha,...,a_{p}+\alpha\\
b_{1}+\alpha,...,b_{q}+\alpha
\end{array}\biggr|\; z\right]
\end{equation}

\noindent 
Among the indefinite and definite integrals of the Meijer G-function 
one has the following:

\begin{equation}
\int G_{p,q}^{m,n}\left[\begin{array}{c}
a_{1},...,a_{p}\\
b_{1},...,b_{q}
\end{array}\biggr|\;\alpha z\right]z^{\alpha-1}dz=
z^{\alpha}G_{p,q}^{m,n}\left[\begin{array}{c}
1-\alpha,a_{1},...,a_{p}\\
b_{1},...,b_{q},-\alpha
\end{array}\biggr|\; z\right]\label{eq:indefinite-integration-meijer}
\end{equation}

\begin{eqnarray}
& \int_{0}^{\infty}G_{p,q}^{m,n} & 
\left[\begin{array}{c}
a_{1},...,a_{p}\\
b_{1},...,b_{q}
\end{array}\biggr|\;\beta z\right]z^{\alpha-1}dz \nonumber \\ 
& = & 
\frac{{\textstyle 
\prod_{j=1}^{m}\Gamma\left(b_{j}+\alpha\right)\prod_{j=1}^{n}
\Gamma\left(1-a_{j}-\alpha\right)}}{{\textstyle \prod_{j=m+1}^{q}
\Gamma\left(1-b_{j}-\alpha\right)\prod_{j=n+1}^{p}
\Gamma\left(a_{j}+\alpha\right)}}\beta^{-\alpha}
\end{eqnarray}

\noindent 
A short list of relations between the Meijer G-function
and some elementary and special functions is:

\begin{equation}
G_{0,1}^{1,0}\left[\begin{array}{c}
-\\
b
\end{array}\biggr|\; z\right]=\exp\left(-z\right)z^{b}
\label{eq:exponetial-meijer}
\end{equation}

\begin{equation}
G_{0,2}^{2,0}\left[\begin{array}{c}
-\\
b_{1},b_{2}
\end{array}\biggr|\; z\right]=2z^{\frac{1}{2}\left(b_{1}+b_{2}\right)}
K_{b_{1}-b_{2}}\left(2\sqrt{z}\right) \label{eq:bessel_second-meijer}
\end{equation}

\noindent 
A more complete list can found in \citet{bateman1953higher}
and the Wolfram Functions 
Site\footnote{http://functions.wolfram.com/HypergeometricFunctions/MeijerG/}.

\noindent 
The Fox H-function is a generalization of the Meijer G-function
introduced by \citet{1961TAMS...98..395}. It is defined in terms
of an Mellin inverse transform:

\[
H_{p,q}^{m,n}\left[\begin{array}{c}
(\mathbf{a},\mathbf{A})\\
(\mathbf{b},\mathbf{B})
\end{array}\biggr|\; z\right]\equiv H_{p,q}^{m,n}\left[\begin{array}{c}
(a_{1},\, A_{1}),...,(a_{p},\, A_{p})\\
(b_{1},\, B_{1}),...,(b_{q},\, B_{q})
\end{array}\biggr|\; z\right]
\]

\begin{equation}
=\frac{1}{2\pi i}\int_{_{\mathit{C}}}\frac{{\textstyle \prod_{j=1}^{m}
\Gamma\left(b_{j}+B_{j}s\right)\prod_{j=1}^{n}
\Gamma\left(1-a_{j}-A_{j}s\right)}}{{\textstyle \prod_{j=m+1}^{q}
\Gamma\left(1-b_{j}-B_{j}s\right)\prod_{j=n+1}^{p}
\Gamma\left(a_{j}+A_{j}s\right)}}z^{-s}ds\label{eq:Fox-H-1}
\end{equation}

\noindent 
More properties and applications of the Fox H-function can
be found in 
\citep{085226559X,srivastava1982h,9810206909,0415299160,mathai2009h}.

\end{document}